\documentclass[journal]{IEEEtran}

\usepackage{cite}
\usepackage{amsmath,amssymb}
\usepackage{booktabs}
\usepackage{array}
\usepackage{multirow}
\usepackage{url}
\usepackage{balance}
\usepackage{microtype}
\usepackage{graphicx}

\begin{document}

\title{Dense and Query-Set Prediction for J-Peak Detection in Pillow-Based Ballistocardiography}

\author{Shengwei~Guo and Guobing~Sun%
\thanks{Shengwei Guo and Guobing Sun are with the Key Laboratory of Information Fusion Estimation and Detection, School of Electronic Engineering, Heilongjiang University, Harbin, China. Corresponding author: Guobing Sun (e-mail: sunguobing@hlju.edu.cn).}%
}

\markboth{IEEE Signal Processing Letters}{Guo and Sun: Dense and Query-Set Prediction for Pillow-BCG J-Peaks}
\maketitle

\begin{abstract}
Pillow-based ballistocardiography (BCG) enables unobtrusive cardiac monitoring, but J-peak detection is commonly learned as dense sequence labeling although the desired output is a sparse event set. This letter asks a narrower question: how do dense and query-set outputs differ when the data, encoder, validation protocol, and event evaluator are controlled? We formulate one-dimensional point-set prediction with 64 learned queries and Hungarian assignment, and compare it with a shared-backbone dense Transformer and a U-Net--BiLSTM. Evaluation uses five-subject leave-one-subject-out testing, three independently seeded runs, validation-only postprocessing selection, and strict one-to-one peak association. The dense Transformer attains the highest subject-wise pooled F1 ($0.786\pm0.105$) and precision ($0.817\pm0.089$), whereas U-Net--BiLSTM obtains $0.780\pm0.094$ F1. Set+DN reaches $0.764\pm0.112$ F1 but the lowest beat-count error ($0.862\pm0.584$ beats/epoch), compared with $2.790\pm1.356$ for the dense Transformer. DN changes set-model F1 by only $+0.004$. The results identify distinct event-accuracy and count-fidelity operating points; they do not establish universal superiority of either output formulation.
\end{abstract}


\section{Introduction}
Ballistocardiography (BCG) measures the body micro-motion caused by cardiac ejection and can be acquired unobtrusively from beds or pillows~\cite{inan2015review,sadek2019review}. Such sensing is attractive for long-duration sleep monitoring, but the BCG waveform varies with posture, body--sensor coupling, respiration, morphology, and motion~\cite{li2024dataset,qiu2025dataset}. Reliable J-peak detection is therefore a central signal-processing step. Extra peaks inflate fixed-window beat counts and introduce short inter-beat intervals (IBIs); missed peaks reduce counts and merge adjacent intervals; localization error perturbs otherwise correct IBIs. These error pathways motivate reporting event detection, timing, interval, and count metrics rather than a single accuracy number.

Classical BCG detectors use morphology, interval, template, or envelope assumptions~\cite{bruser2011adaptive,mora2020accurate}. Deep systems increasingly predict a dense sample-wise confidence sequence and recover peaks by thresholding and temporal suppression, including U-Net, U-Net++, U-Net--BiLSTM, and surrogate-event models~\cite{cathelain2020unet,zhou2021unetpp,mai2022unetbilstm,schranz2024surrogate}. Dense prediction is effective, but the learned representation and the reported event set are linked through a peak-extraction rule. Event-based biomedical models instead learn centers, durations, or unordered event sets~\cite{seeuws2024event}. DETR introduced finite learned queries and permutation-invariant assignment~\cite{carion2020detr}; related query formulations have since been applied to EEG segmentation and asynchronous EEG event detection~\cite{wolf2022detrtime,levy2026dance}.

\begin{table}[!t]
	\caption{Closest output formulations. Query models support a variable reported count subject to a finite cap $K$.}
	\label{tab:related}
	\centering
	\scriptsize
	\begin{tabular}{@{}p{0.275\columnwidth}p{0.215\columnwidth}p{0.28\columnwidth}p{0.14\columnwidth}@{}}
		\toprule
		Work & Signal/task & Learned output & Assignment \\
		\midrule
		Dense BCG~\cite{cathelain2020unet,mai2022unetbilstm,schranz2024surrogate} & BCG beats & sample confidence map & none \\
		Seeuws \emph{et al.}~\cite{seeuws2024event} & biomedical events & center/duration maps & none \\
		DETRtime~\cite{wolf2022detrtime} & EEG segments & segment queries & Hungarian \\
		DANCE~\cite{levy2026dance} & EEG events/classes & event queries & set matching \\
		This work & BCG J-peaks & point queries + dense cue & Hungarian \\
		\bottomrule
	\end{tabular}
\end{table}

Table~\ref{tab:related} makes the scope explicit: queries and Hungarian matching are established ideas. The unresolved BCG question is whether a point-set output offers a useful operating point relative to dense output under a subject-disjoint and representation-controlled comparison. We address it by: 1) defining the finite-capacity BCG point-set objective and disclosing the complete inference paths; 2) comparing a query decoder with a dense head after an identical convolutional backbone and Transformer encoder, alongside a strong U-Net--BiLSTM reference; and 3) evaluating all four configurations over five held-out subjects and three seeds with strict one-to-one metrics. The contribution is therefore a controlled BCG formulation study and its error-profile finding, not a claim that set prediction is intrinsically superior.

\section{Problem Formulation and Models}
\subsection{Point-set formulation}
Let $\mathbf{x}\in\mathbb{R}^{T}$ be a z-normalized BCG epoch and $\mathcal{Y}=\{t_j\}_{j=1}^{N}$ its J-peak sample indices, with normalized coordinates $u_j=t_j/(T-1)$. A query detector emits
\begin{equation}
\widehat{\mathcal{Y}}_K=\{(p_k,\hat u_k)\}_{k=1}^{K},
\end{equation}
where $p_k$ is the peak probability and $\hat u_k\in[0,1]$. The finite budget imposes $|\widehat{\mathcal{Y}}|\le K$ and a recall ceiling when $N>K$. Here $K=64$, while the retained epochs contain 15--42 annotated peaks, so no observed epoch exceeds this cap.

\begin{figure*}[!t]
	\centering
	\includegraphics[width=0.48\textwidth]{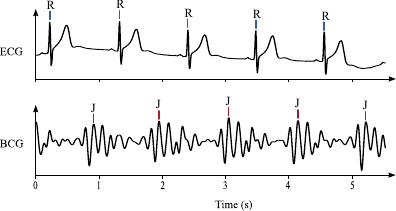}
	\includegraphics[width=0.48\textwidth]{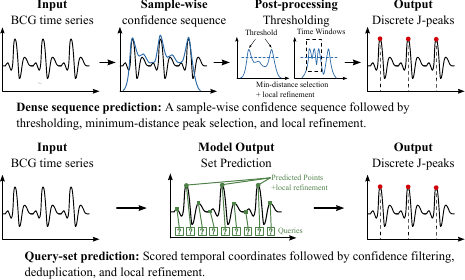}
	\caption{ECG-assisted annotation and the compared output formulations.
		(a) Synchronized ECG R-peaks provide temporal references for manual
		BCG J-peak annotation; the illustrated R--J relation is schematic and
		does not assume a fixed delay.
		(b) Given the same BCG epoch and shared backbone--Transformer encoder,
		dense prediction produces a length-$T$ sample-wise confidence sequence,
		followed by validation-selected thresholding and minimum-distance peak
		extraction. Query-set prediction produces $K$ scored temporal
		coordinates, followed by confidence filtering, duplicate suppression,
		and local refinement. Both branches show their complete principal
		inference stages; the diagram does not imply a postprocessing-free or
		causal advantage for either formulation.}
	\label{fig:pipelines}
\end{figure*}

For target $j$ and query $k$, Hungarian assignment minimizes
\begin{equation}
C_{kj}=-p_k+c_m|\hat u_k-u_j|,\qquad c_m=15.
\end{equation}
Matched queries are labeled as peaks and unmatched queries as no-event. The set loss is
\begin{equation}
\mathcal{L}_{\rm set}=\mathcal{L}_{\rm cls}+15\mathcal{L}_{1},
\end{equation}
with no-event class weight 0.02. Both set configurations also use an auxiliary dense heat-map loss $2\mathcal{L}_{\rm heat}$. Set+DN additionally injects 20 target-conditioned denoising queries during training, with coordinate perturbation 0.02 and DN coordinate weight 5~\cite{li2022dndetr}; Set w/o DN omits only this term. The implemented DN branch tiles or truncates the time-ordered target list to 20 queries; when $N>20$, only the first 20 targets are used. This implementation-specific construction does not alter $K$ or inference.

The roles of $K$, $c_m$, and DN are distinct but coupled through optimization. $K$ fixes the maximum number of represented events; once it exceeds the epoch cardinality, excess queries are trained as no-event. The matching cost $c_m$ determines how strongly temporal proximity influences which query claims a target, whereas the coordinate-loss weight determines how strongly the matched coordinate is optimized. DN supplies perturbed target-centered queries and therefore operates on the same normalized coordinate scale. We fix these values for the controlled comparison and do not infer hyperparameter optimality from the final model ranking.

\subsection{Controlled dense and set pipelines}
All Transformer configurations use the same four-stage stride-2 convolutional backbone ($d=128$), sinusoidal position encoding, and two-layer Transformer encoder with eight heads and feed-forward width 256. The \emph{Dense Transformer} upsamples encoder memory through a convolutional dense head. The \emph{Set} models add a two-layer Transformer decoder, 64 learned queries, classification/coordinate heads, and the auxiliary heat head. This shared encoder reduces, but does not eliminate, formulation confounding because the output head and query decoder remain different. U-Net--BiLSTM uses the same convolutional backbone, a four-stage U-Net decoder, and one bidirectional LSTM layer with 128 hidden units per direction.

Figure~\ref{fig:pipelines} makes the representation contrast explicit. The reported results compare complete pipelines: dense probabilities are thresholded and suppressed, then snapped within $\pm30$ samples to a local maximum of a nine-sample-smoothed BCG signal. Set inference applies confidence filtering with a top-five fallback, confidence-ranked deduplication, $\pm20$-sample auxiliary-heat refinement, a second minimum-distance deduplication, the same snapping, and final 10-sample deduplication. Thresholds and minimum distances are selected using validation data only and frozen for the held-out subject. Thus, the comparison is neither query-only nor postprocessing-free.

\section{Evaluation Protocol}
The cohort contains five healthy male participants recorded over eight nights with synchronized pillow BCG and ECG reference timing.
 The study was approved before data collection by the Beijing Anzhen Hospital Medical Ethics Committee (Approval No.~KS2023009), and all participants provided informed consent prior to participation.
 After the supplied quality-control index, 3369 non-overlapping epochs remain. Each epoch contains $T=4000$ samples at 133~Hz ($\approx30.1$~s) and is z-normalized independently.

We perform five outer leave-one-subject-out (LOSO) folds. In each fold, one subject is reserved for testing; epochs from the other four subjects are randomly divided 80/20 into training and validation sets. The same seed/fold partition is used for every model. The seeds 13, 42, and 2025 control both stochastic training and this inner epoch split. Hyperparameters and postprocessing are selected by equal-weight mean validation F1 across the four non-test subjects; the test subject is never used for selection. Models are trained for 200 epochs with batch size 32, AdamW, weight decay $10^{-2}$, and a one-cycle schedule with maximum learning rate $3\times10^{-4}$.

The confidence grid spans $10^{-4}$--0.9 for dense models and $10^{-4}$--0.999 for set models. It was expanded using validation data until no selected threshold remained at a grid boundary. Four dense runs selected the predefined 30-sample minimum-distance floor; their validation advantage over the nearest interior value was at most $3.91\times10^{-4}$. All 60 checkpoints loaded without missing or unexpected parameters, and checkpoint/split hashes were recorded before evaluation.

A predicted and reference peak are associated one-to-one by optimal assignment only when their distance is at most 10 samples (75.2~ms). TP, FP, and FN are pooled over all epochs within each held-out subject to compute precision, recall, and F1. Localization MAE pools absolute errors over matched pairs. Matched-IBI MAE compares intervals between consecutive matched pairs, and count MAE is
\begin{equation}
\operatorname{CMAE}=\frac{1}{M}\sum_{m=1}^{M}\left|\,|\widehat{\mathcal{Y}}_m|-|\mathcal{Y}_m|\,\right|.
\end{equation}
For each subject, metrics are first averaged over the three seeds; Table~\ref{tab:results} reports mean $\pm$ sample SD across five subjects. Paired differences use these five subject-level values. We report exploratory 20,000-resample subject-bootstrap intervals and exact two-sided sign-flip values; with $n=5$, the minimum attainable two-sided value is 0.0625, so interpretation emphasizes effect sizes and direction rather than binary significance. Evaluation code, splits, checkpoint hashes, and result manifests accompany the submission.

\section{Results and Discussion}
\begin{table*}[!t]
\caption{Three-seed LOSO results. Subject columns are seed-averaged pooled F1. Summary metrics are mean $\pm$ sample SD across the five held-out subjects. Loc. is pooled matched-pair localization MAE; IBI is matched-IBI MAE; Count is beats/epoch. Best mean in each column is bold.}
\label{tab:results}
\centering
\scriptsize
\begin{tabular}{lcccccc}
\toprule
Model & S1 & S2 & S3 & S4 & S5 & F1 \\
\midrule
Dense Transformer & \textbf{.824} & .686 & .672 & \textbf{.920} & .829 & \textbf{.786$\pm$.105} \\
U-Net--BiLSTM & .807 & \textbf{.691} & \textbf{.677} & .898 & .825 & .780$\pm$.094 \\
Set+DN & .797 & .637 & .657 & .892 & \textbf{.838} & .764$\pm$.112 \\
Set w/o DN & .804 & .634 & .639 & .893 & .830 & .760$\pm$.117 \\
\bottomrule
\end{tabular}

\vspace{2pt}
\begin{tabular}{lcccccc}
\toprule
Model & Precision & Recall & Loc. (ms) & IBI (ms) & Count MAE & Params (M) \\
\midrule
Dense Transformer & \textbf{.817$\pm$.089} & .759$\pm$.119 & \textbf{1.763$\pm$.291} & 2.514$\pm$.275 & 2.790$\pm$1.356 & \textbf{0.809} \\
U-Net--BiLSTM & .791$\pm$.086 & \textbf{.770$\pm$.102} & 1.779$\pm$.299 & \textbf{2.483$\pm$.255} & 1.381$\pm$.980 & 1.664 \\
Set+DN & .767$\pm$.107 & .762$\pm$.118 & 1.781$\pm$.305 & 2.572$\pm$.288 & \textbf{0.862$\pm$.584} & 1.149 \\
Set w/o DN & .763$\pm$.113 & .757$\pm$.122 & 1.780$\pm$.309 & 2.566$\pm$.296 & 0.943$\pm$.574 & 1.149 \\
\bottomrule
\end{tabular}
\end{table*}

\begin{table}[!t]
\caption{Paired subject-level effects. DT denotes Dense Transformer and Set denotes Set w/o DN. $\Delta=A-B$; negative $\Delta$Count MAE favors $A$. CIs are exploratory bootstrap intervals.}
\label{tab:paired}
\centering
\scriptsize
\begin{tabular}{@{}p{0.29\columnwidth}rrr@{}}
\toprule
Comparison ($A$ vs. $B$) & $\Delta$F1 [95\% CI] & A$>$B (F1) & $\Delta$Count MAE \\
\midrule
DT vs. U-Net & .0066 [$-.0032$, .0164] & 3/5 & 1.409 \\
DT vs. Set & .0262 [.0103, .0414] & 4/5 & 1.847 \\
Set vs. U-Net & $-.0196$ [$-.0423$, $-.0004$] & 1/5 & $-.439$ \\
Set+DN vs. Set & .0044 [$-.0027$, .0125] & 3/5 & $-.080$ \\
\bottomrule
\end{tabular}
\end{table}

\subsection{Event accuracy versus count fidelity}
Dense Transformer has the highest mean event F1 and precision while using the fewest parameters. Relative to U-Net--BiLSTM, its paired F1 difference is $+0.0066$ (Table~\ref{tab:paired}; exact sign-flip $p=0.4375$). The two dense models are close in mean F1, but Dense Transformer has higher precision in all five subjects, whereas U-Net--BiLSTM has the highest mean recall. Dense Transformer uses 51.4\% fewer parameters than U-Net--BiLSTM; this is a model-size result, not a latency or memory measurement.

The closest output-formulation control is Dense Transformer versus Set w/o DN because they share the convolutional backbone and Transformer encoder. Dense output improves F1 by 0.0262 and precision by 0.0540, with F1 higher in four subjects and precision higher in all five. Conversely, Set w/o DN reduces count MAE by 1.847 beats/epoch in every subject. Against U-Net--BiLSTM, it lowers count MAE by 0.439 beats/epoch in all five subjects while its F1 is lower by 0.0196. Set+DN reduces count MAE by 37.6\% relative to U-Net--BiLSTM and by 69.1\% relative to Dense Transformer. These observations reveal an event-accuracy--cardinality trade-off rather than a uniformly better representation.

The lower set-model count error is consistent with one-to-one assignment and explicit no-event queries discouraging duplicate events. It is not identified as a causal effect of set output because the decoder and inference path also differ. Importantly, localization errors are nearly identical (1.76--1.78~ms), and matched-IBI errors differ by less than 0.1~ms. The principal distinction is therefore not the timing of successfully associated beats; it is the balance between event precision and the number of reported beats.

\subsection{Subject heterogeneity and DN}
The same subjects are difficult for all models: F1 is lowest for Subjects 2 and 3 and highest for Subject 4. Nevertheless, the best method varies by subject: Dense Transformer leads Subjects 1 and 4, U-Net--BiLSTM Subjects 2 and 3, and Set+DN Subject 5. The subject-wise SDs of 0.094--0.117 are comparable to, or larger than, the between-model mean F1 gaps. This pattern is why the paper reports every held-out subject and avoids population-level ranking from five participants.

Set+DN changes F1 by only $+0.0044$ relative to Set w/o DN ($p=0.3750$) and lowers count MAE by 0.080 beats/epoch in four of five subjects. Differences in localization and matched-IBI error are negligible. DN is therefore a modest training variant rather than a demonstrated source of general accuracy or robustness. The observed small change also indicates that the main dense--set contrast is not explained by DN alone.

\subsection{Downstream meaning and error pathways}
For a fixed 30.1-s epoch, count error maps directly to heart-rate error through multiplication by approximately $60/30.1\approx1.995$ beats/min per count. Thus, the mean count MAEs correspond to about 5.56, 2.75, 1.72, and 1.88 beats/min for Dense Transformer, U-Net--BiLSTM, Set+DN, and Set w/o DN, respectively. This conversion clarifies practical scale but is not a separate heart-rate experiment.

\begin{table}[!t]
\caption{How peak errors propagate to the reported metrics and downstream beat sequences.}
\label{tab:errorpaths}
\centering
\scriptsize
\begin{tabular}{@{}p{0.16\columnwidth}p{0.25\columnwidth}p{0.12\columnwidth}p{0.34\columnwidth}@{}}
\toprule
Outcome & Event metric & Count & Interval consequence \\
\midrule
extra peak & FP; precision $\downarrow$ & $+1$ & one interval split, often short \\
missed peak & FN; recall $\downarrow$ & $-1$ & adjacent intervals merged \\
matched shift & TP within tolerance & unchanged & localization and matched IBI perturbed \\
artifact burst & multiple FP/FN & variable & irregular local sequence \\
\bottomrule
\end{tabular}
\end{table}

Table~\ref{tab:errorpaths} explains why F1 and count MAE need not rank models identically. Subject-wise pooled F1 normalizes TP by FP and FN over an entire held-out subject, whereas count MAE gives each epoch equal weight and measures only cardinality. A model may therefore improve global precision by suppressing uncertain beats yet retain a sizable per-epoch count error, or report a more accurate number of beats while placing enough individual events incorrectly to reduce F1. Matched-IBI MAE is deliberately conditional: it compares intervals formed by consecutive matched pairs. It does not replace penalties for missing or extra beats, which remain visible in F1 and count MAE.

For downstream use, false positives can create spuriously short intervals and false negatives can create long merged intervals. These errors can affect time-domain HRV features even when the mean heart rate over a long window is acceptable~\cite{taskforce1996hrv,shaffer2017hrv}. Conversely, the 1.76--1.78-ms matched localization MAEs are small relative to the 7.52-ms sampling period and the 75.2-ms association tolerance. The present results therefore support a statement about detection and beat-sequence fidelity, not validated HRV, sleep-staging, or clinical performance.

\subsection{What is identified by the controlled comparison}
The shared-backbone Dense Transformer versus Set w/o DN contrast is the most direct architectural comparison in this study: both receive the same input, use the same downsampling backbone and Transformer encoder, and are trained/tested on identical seed/fold partitions. The contrast removes encoder capacity as an explanation for the result, but it does not isolate a single mechanism. Query decoding, no-event classification, auxiliary heat supervision, and the postprocessing path all belong to the set pipeline. Accordingly, we describe associations between complete configurations rather than a causal benefit of a particular component.

The U-Net--BiLSTM reference serves a different purpose. It tests whether the compact shared-encoder Transformer family is competitive with a recurrent dense detector commonly used for BCG. Dense Transformer achieves a slightly higher mean F1 with roughly half the trainable parameters, while U-Net--BiLSTM preserves slightly higher recall and lower count error. The result balances accuracy and model size, but parameter count alone is not computational efficiency: attention, recurrent execution, memory access, and device implementation can change latency. FLOPs, energy, and wall-clock deployment measurements are outside the present evidence.

The DN comparison is more narrowly interpretable because the architecture and inference path are identical. Its small F1 and count changes indicate that DN is not the source of the main dense--set difference in this setting. However, the experiment does not establish that DN is generally unnecessary: its effect can depend on query budget, coordinate perturbation, matching scale, data volume, and training duration. We therefore retain the exact configuration and report the observed effect without extrapolating beyond this cohort.

\subsection{Assumptions, failure boundaries, and reproducibility}
The set formulation has an explicit capacity boundary $N\le K$. It is inactive for the retained data but could fail under tachycardia, longer windows, or a different sampling/segmentation policy. Dense prediction has no query cap, but its output count is sensitive to calibration, local maxima, and minimum-distance suppression. The set pipeline is also not free of heuristics: confidence filtering, fallback selection, heat refinement, deduplication, and BCG snapping can affect the final set. Epoch boundaries may additionally truncate a cardiac cycle; because all models use the same non-overlapping epochs and evaluator, this is a shared benchmark condition rather than a resolved physiological issue.

The quality-control index removes unsuitable epochs before modeling. Consequently, the benchmark characterizes retained natural-sleep BCG rather than robustness to severe movement or sensor detachment. Subject 2 and Subject 3 remain difficult across all models, suggesting that normal between-subject morphology and coupling variation already dominate the small model-to-model mean gaps. The current five-subject LOSO protocol supports held-out-identity evaluation, but not demographic coverage or confidence in rare failure modes.

The archive contains 15 train/validation/test split files, 60 checkpoint hashes, validation sweeps, state-dictionary integrity checks, one-to-one association tests, and the 60 subject--seed rows used to regenerate the numerical results tables. The code is the best available reconstruction associated with the archived checkpoints; these records improve traceability but do not substitute for external validation.

The study is bounded to five healthy men, one pillow sensor, and no external or pathological cohort. The inner 80/20 split is epoch-level within the four non-test identities, although the test subject is isolated. The $K=64$ check applies only to retained 15--42-peak epochs, and the 20-target DN truncation is implementation-specific. These limits preclude causal, clinical, or demographic generalization.

\section{Conclusion}
Under three-seed five-subject LOSO evaluation, Dense Transformer provides the highest event F1/precision and fewest parameters, whereas query-set output markedly reduces beat-count error with nearly unchanged matched timing. DN changes the set result only slightly. These complete pipelines occupy different event-accuracy/count-fidelity operating points on this benchmark; neither formulation is inherently superior. Broader claims require larger, diverse, artifact-rich, externally validated cohorts.

\clearpage
\balance

\clearpage
\begin{IEEEbiography}
	[{\includegraphics[
		width=1.0in,
		height=1.2in,
		clip,
		keepaspectratio
		]{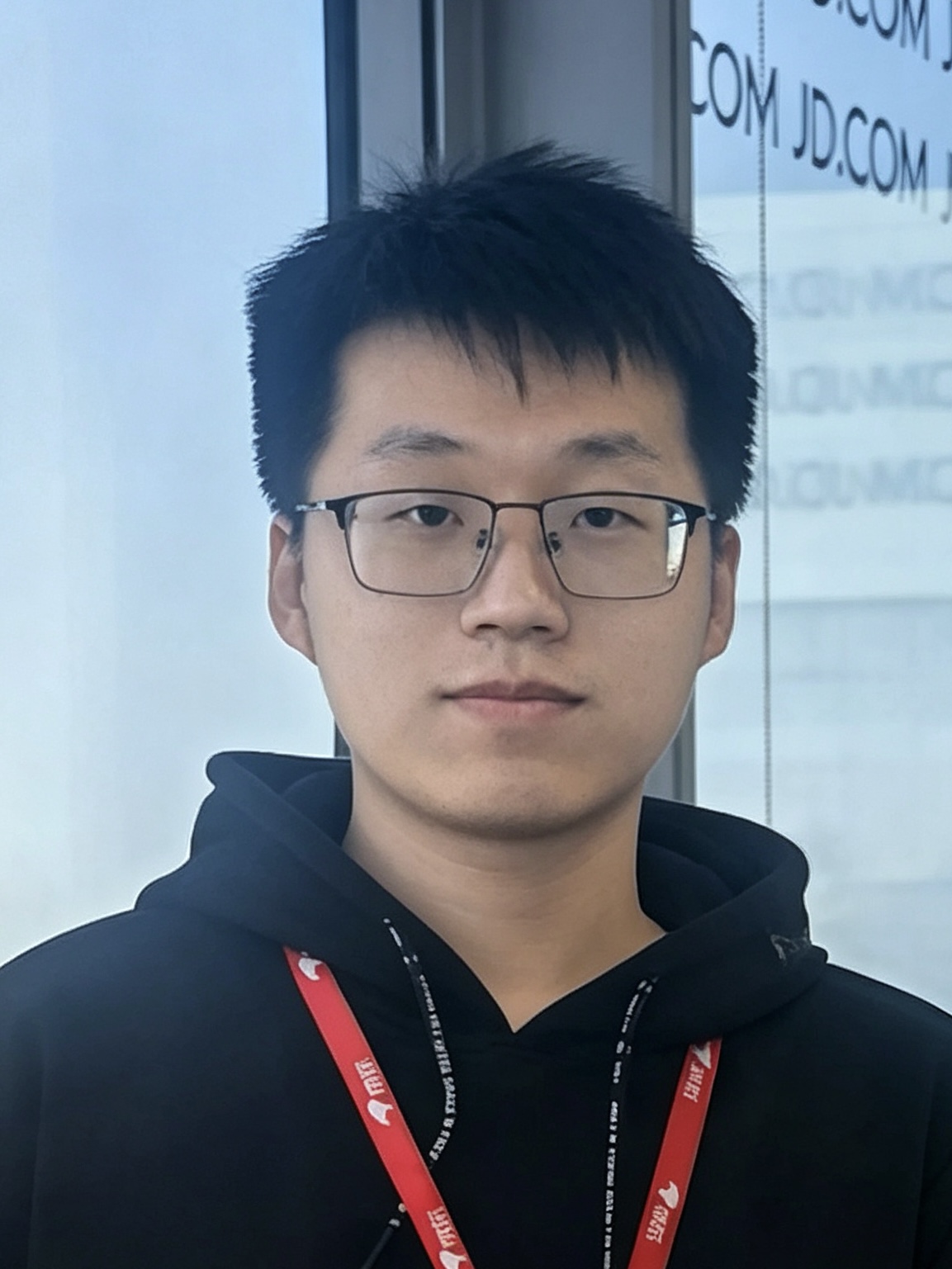}}]
	{Shengwei Guo}
	received the B.E. degree in automation and the M.Eng. degree in electronic information from Heilongjiang University, Harbin, China, in 2020 and 2024, respectively. He is currently with the Key Laboratory of Information Fusion Estimation and Detection, School of Electronic Engineering, Heilongjiang University. His research interests include automatic sleep staging, neurophysiological signal processing, computer vision, and robotics.
\end{IEEEbiography}


\begin{IEEEbiography}
	[{\includegraphics[
		width=1.0in,
		height=1.2in,
		clip,
		keepaspectratio
		]{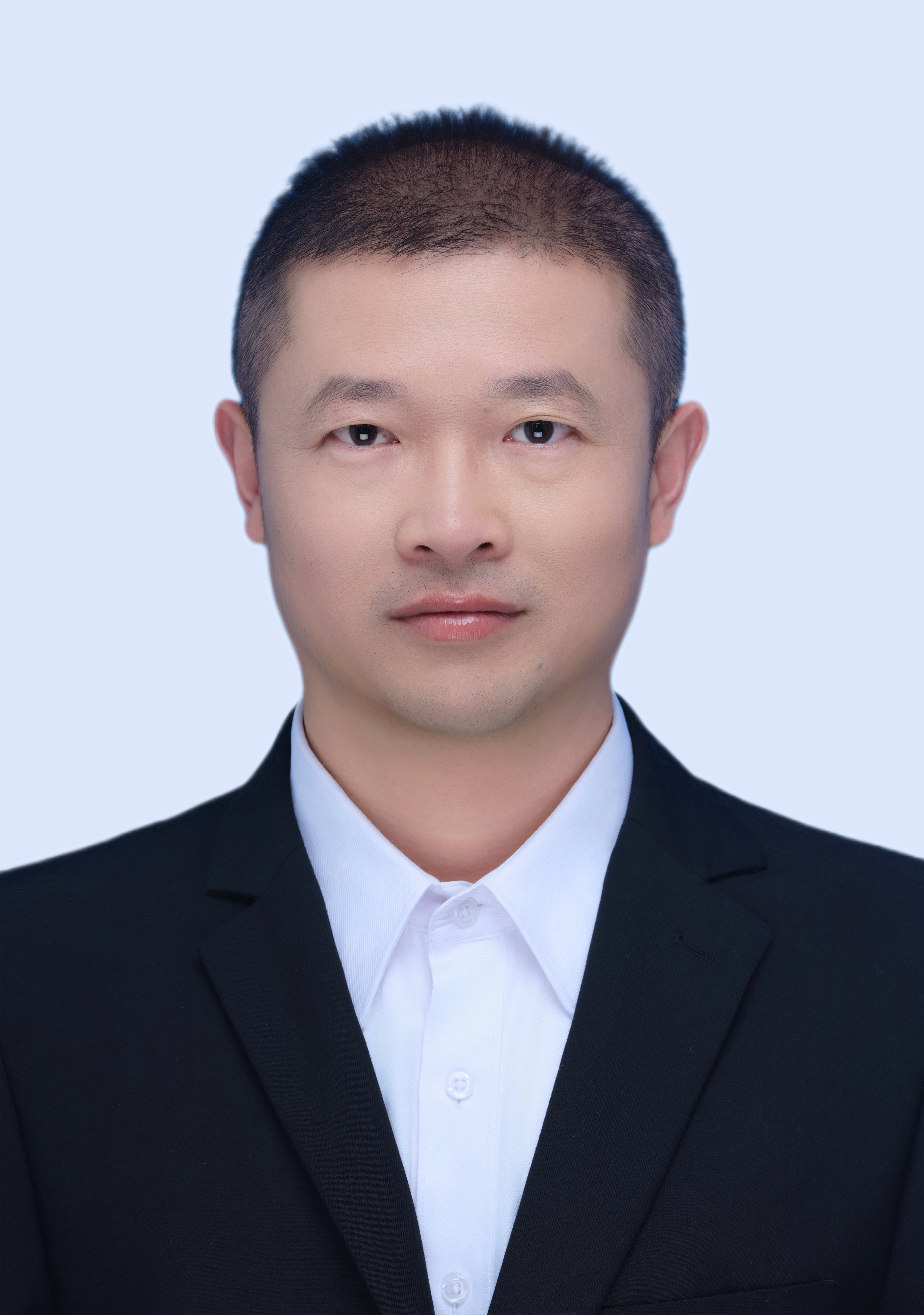}}]
	{Guobing Sun}
	received the B.E. and Ph.D. degrees from Harbin Institute of Technology, Harbin, China, in 2002 and 2009, respectively. He is currently an Associate Professor with the Key Laboratory of Information Fusion Estimation and Detection, School of Electronic Engineering, Heilongjiang University, Harbin, China. His research interests include biomedical signal processing, machine learning, computer vision, and intelligent control.
\end{IEEEbiography}

\end{document}